\documentclass[runningheads]{llncs}

\usepackage{multirow}
\usepackage{booktabs}
\usepackage{comment}
\usepackage{amsmath}
\usepackage{amsbsy}
\usepackage{subcaption}

\usepackage{graphicx}
\usepackage{enumitem}

\begin{document}

\title{Should I Stay or Should I Go: Predicting Changes in Cluster Membership}

\author{Evangelia Tsoukanara\inst{1} \and Georgia Koloniari\inst{1} \and Evaggelia Pitoura\inst{2}}

\authorrunning{Tsoukanara et al.}

\institute{Department of Applied Informatics, University of Macedonia, 
Thessaloniki, Greece \email{\{etsoukanara, gkoloniari\}@uom.edu.gr}\\
\and
Computer Science \& Engineering, University of Ioannina, Ioannina, Greece
\email{pitoura@cse.uoi.gr}}

\maketitle

\begin{abstract}
Most research on predicting community evolution 
focuses on changes in the states of communities. Instead, we focus on
individual nodes and define the novel problem of predicting
whether a specific node stays in the same cluster, moves to another cluster or
drops out of the network. We explore variations of the problem and propose
appropriate classification features based on local and global node measures. Motivated by the prevalence of machine learning approaches based on embeddings, we also introduce efficiently computed distance-based features using appropriate node embeddings. 
In addition, we consider chains of features to capture the history of the nodes.
Our experimental results depict the complexity of the different formulations of the problem and the suitability of the selected features and chain lengths.

\end{abstract}


\section{Introduction}
For the problem of community evolution, communities are monitored across time and their properties studied to attain useful conclusions. Such conclusions can then be exploited so as to predict community changes through time. Most works model community evolution through a predefined set of events, such as community growth or shrinkage, community merging or splitting and so on \cite{Gliwa13,lhan15,Pavlopoulou17,Saganowski15}. The problem is then modeled as a classification problem, where given a community history, the next event in its evolution is predicted. 

 However, in many applications, it is important not only to predict the behavior of a community but also of individual community members. Let us consider customers that are connected via the common products they buy. These customers can be clustered according to the companies and products they prefer. Focusing on individuals in such clusters, makes it possible for companies to identify and reward loyal customers (community members) or take preemptive measures to change the behavior of the ones that seem less dedicated.
 

To this end, we introduce a novel community evolution prediction problem defined at node level. For each individual node, there are three possible events: the node may (a) stay in the same cluster, (b) move to a different cluster, or (c) drop out of the network.  We study variations of the problem by considering combinations and subsets of the possible evolution events, and evaluate the use of different classification features for solving them. Firstly, we consider classic features based on popular node measures. Motivated by the advances of machine learning methods for community detection that use node embeddings, we also propose features based on distances between such embeddings. In particular, we deploy the ComE \cite{ComE17} approach that provides both node and community embeddings. Features of both methods, defined at cluster and out of cluster or network level, are combined into chains, modeling the evolution history of the nodes through time \cite{Saganowski15}.  Our  experimental results show that the problems we defined are not trivial, and that the proposed distance-based embeddings perform almost as well as the classic ones, while being computed much more efficiently.

The rest of the paper is structured as follows. Section 2 briefly describes related work. In Section 3, we formulate the novel problem and its variations. Section 4 presents our classification features and their modeling into chains. Section 5 includes our experimental results, while Section 6 concludes.
\section{Related Work}
We discuss related research, first, on community evolution and then, on node and graph embeddings.

\noindent\textbf{Community Evolution.}
After discovering communities in evolving networks \cite{Rosetti18}, their properties are studied by mapping corresponding communities through successive network snapshots \cite{Aynaud13}. Community evolution is assessed with measures such as its growth and disappearance rate \cite{Toyoda03}, or its life expectancy \cite{Palla07}.

For predicting community evolution \cite{Gliwa13,Pavlopoulou17,Saganowski15}, events such as community growth, shrinkage, merging and splitting are defined. The problem is modeled as a classification problem in which features based on the structural properties of communities are exploited, and given the history of a community the next event in the community's evolution is predicted.
In our work, instead of communities, we focus on nodes, and introduce a new problem aiming at predicting changes in the nodes' cluster membership through time.

\noindent\textbf{Node Embeddings.}
An embedding is the transformation of a high-dimensional space to a low-dimension vector. For graphs, the focus is mostly on node embeddings that preserve network structure.
Deepwalk \cite{deepwalk14} learns node embeddings that capture second-order proximity (i.e., proximity between shared neighbors) by simulating short random walks and applying the Skip-gram algorithm. LINE \cite{line15} employs edge-sampling, and to preserve both first-order (i.e., ties between neighbors) and second-order proximity, it is first trained separately and then the two resulting embeddings are concatenated. Node2vec \cite{node2vec16} extends Deepwalk by generating biased random walks to explore diverse node neighborhoods. Finally, GraRep \cite{grarep15} and HOPE \cite{hope16} derive embeddings that capture high-order proximity.

Besides, link prediction, node classification and visualization, node and graph embeddings are also used for community detection \cite{Kozdoba15,Tian14}. ComE (Community Embedding) \cite{ComE17} is a framework that jointly solves both community detection, and learning node and community embeddings. The intuition is that node embeddings that capture community-aware proximity, can assist community detection, while community embeddings can in turn improve node embeddings. In our work, we deploy ComE and explore whether node and community embeddings can be used to derive predictive features.
\section{Problem Formulation}

A social network is often represented as a graph $G=(V,E)$, where $V$ is the set of nodes (vertices) and $E$ is the set of edges. A temporal social network is a network that changes over time and is represented as a sequence of graphs $\{G_1, G_2, \dots, G_n\}$, where $G_i=(V_i, E_i)$, represents a snapshot of graph $G$ at time $i$, and $V_i$ and $E_i$ are the node and edge sets at time $i$ respectively. Let $\mathcal{C}_i=\{C^1_i, C^2_i, \dots, C^m_i\}$ be a clustering of $G_i$ consisting of $m$ clusters, such that $C^j_i \cap C^k_i=\emptyset, 1\le j,k \le m, j\neq k$. For two clusterings $\mathcal{C}_i$ and $\mathcal{C}_{i+1}$ at consecutive timeframes $i$ and $i+1$, we assume cluster $C^j_{i+1}$ is the evolution of cluster $C^j_i$. Similarly, for multiple consecutive timeframes, $C^j_1, C^j_2, \dots, C^j_n$ denotes the evolution of cluster $C^j$ in time period $[1, n]$.


Inspired by the idea of community evolution, we study the problem at node level by aiming to predict how node memberships in clusters evolve through time. We discern between different states that a node can have with respect to its cluster membership in the next timeframe, i.e., a node can stay in the same cluster, move to another or drop out of the network. Based on the above, we define our problem as follows.
\begin{definition}[Stay\textbackslash Move\textbackslash Drop ($\mathcal{SMD}$) Problem]
Given a sequence of clusterings $\mathcal{C}_1,\dots, \mathcal{C}_i$, corresponding to a consecutive set of timeframes for a graph $G$, and node $v \in C^j_i$, predict the state of node $v$ regarding its evolution in the next timeframe $i+1$ as state:
\begin{itemize}[noitemsep,topsep=0pt]
\item \textit{stay}, $\mathcal{S}$: node $v$ stays at the same cluster in $i+1$, that is $v \in C^j_{i+1}$,
\item \textit{move}, $\mathcal{M}$: node $v$ moves to another cluster in $i+1$, that is $v \in C^k_{i+1}$, $k\neq j$, and
\item \textit{drop}, $\mathcal{D}$: node $v$ drops out of the network, that is $v \notin V_{i+1}$.
\end{itemize}
\end{definition}
Thus, we define a classification problem with 3 classes, labeled \textit{stay}, \textit{move} and \textit{drop}. Given the history of a node in a given time period, defined as a sequence of distinct timeframes, we predict its class in the next time frame.

If we are only interested in discerning between loyal  cluster members and members likely to leave, we may simplify our problem to a binary classification problem. This first alternative problem, is derived by merging states $move$ and $drop$, in one class $leave$. Thus, the classes are reformed as follows:\\
\textbf{Stay\textbackslash Leave ($\mathcal{SL}$) Problem:} 
the possible node events are reformed as state:
\begin{itemize}[noitemsep,topsep=0pt]
\item \textit{stay}, $\mathcal{S}$: node $v$ stays at the same cluster in $i+1$, that is $v \in C^j_{i+1}$, and
\item \textit{leave}, $\mathcal{L}$: node $v$ does not remain in the same cluster, that is $v \notin C^j_{i+1}$.
\end{itemize}\vspace{0.2mm}


Finally, we omit the third class of the $\mathcal{SMD}$ problem, and only provide predictions for nodes that remain in the network in timeframe $i+1$. For the third variation, we have: \\
\textbf{Stay\textbackslash Move ($\mathcal{SM}$) Problem.} 
\begin{itemize}[noitemsep,topsep=0pt]
\item \textit{stay}, $\mathcal{S}$: node $v$ stays at the same cluster in $i+1$, that is $v \in C^j_{i+1}$, and
\item \textit{move}, $\mathcal{M}$: node $v$ moves to another cluster in $i+1$, that is $v \in C^k_{i+1}$, $k \neq j$.
\end{itemize}

\begin{figure}
\vspace{-0.05in}
\centering
\begin{subfigure}[t]{0.3\textwidth}\centering
\vspace{-0.82in}\includegraphics[scale=0.28]{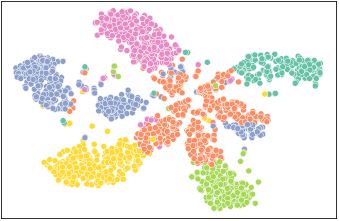}
\vspace{0.175in}
\caption{$T_0$}
\label{fig4.1:sub1}
\end{subfigure}
\begin{subfigure}[t]{0.3\textwidth}\centering
\includegraphics[scale=0.17]{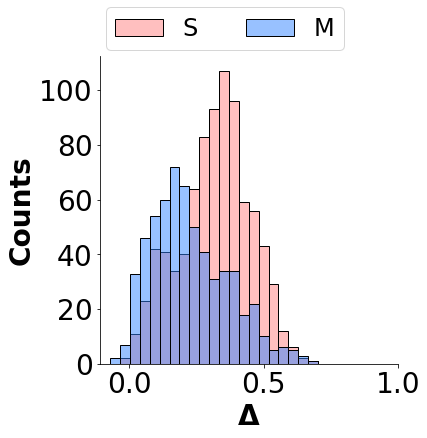}
\caption{\textit{Syntgen} dataset}
\label{fig4.1:sub2}
\end{subfigure}
\begin{subfigure}[t]{0.3\textwidth}\centering
\includegraphics[scale=0.17]{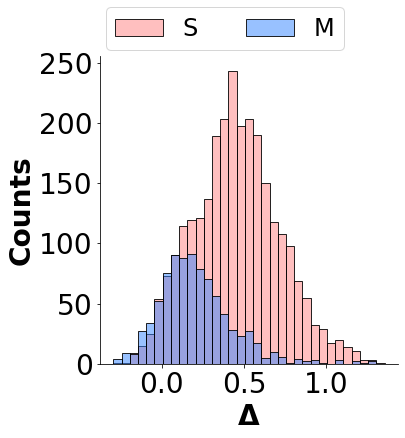}
\caption{\textit{DBLP} dataset}
\label{fig4.1:sub3}
\end{subfigure}
\vspace{-0.05in}
\caption{(a) ComE clustering and $\Delta$ distribution for (b) \textit{Syntgen} and (c) \textit{DBLP}.}
\label{fig:4.1}
\vspace{-0.1in}
\end{figure}
\section{Predictive Features \& Historical Chains}
To solve the classification problems we define, the evaluation and selection of appropriate predictive features is required. We discern between two basic types of features, based on structural network measures that are usually exploited for community evolution predictions, and on distances between embeddings that we propose. As all three problems we introduce are variations of the same classification problem, we define and evaluate the same features for all problems.

\noindent\textbf{Classic Features.}
Classic features, defined on node level, provide information about the structural role of a node in the network. In particular, we select: (a) \textit{degree} measuring the connections of a node, (b) \textit{betweenness}, measuring the number of shortest paths that pass through a node, (c) \textit{closeness}, that measures the distance of a node to all other nodes, and (d) \textit{eigenvector} centrality measuring the influence of a node in the network, defined on cluster and network level.

Aggregated at community level, classic features offer insights for predicting community evolution  \cite{Pavlopoulou17,Saganowski15}, while for individual nodes, such measures also contain information about their evolution tendencies. For instance, an influential central node with high degree is less likely to drop out of the network compared to a low-degree remote node. Similarly focusing on community structure, well-connected core nodes within a community are more likely to stay in their cluster compared to loosely connected border nodes. Thus, we also differentiate between features defined at cluster level (\textit{in}) and at network level (\textit{out}).
 
\noindent\textbf{Embeddings-based Features.}
Since node embeddings are low dimensional vector representations of nodes that capture structural graph properties, we propose defining predictive features based on such embeddings. The ComE \cite{ComE17} framework provides an appropriate solution by solving both community detection and embeddings learning jointly, focusing on deriving embeddings that capture community-aware proximity between nodes. ComE uses as input a graph's edge list and the number of target clusters $k$ and outputs: (i) for each node its embedding along with its community membership and (ii) for each community, which is defined as a multivariate Gaussian distribution, its embedding parameters, that is, a median vector (i.e., the embedding of the mean of the community) and a covariance matrix. Fig. \ref{fig4.1:sub1} depicts the node embeddings for the six clusters detected by ComE for a synthetic dataset generated by the Syntgen \cite{Pereira19} generator for timeframe $T_0$, where we notice that ComE manages to detect reasonable well-separated clusters with similar node embeddings as projected in 2-d.

We define features on cluster and network level, exploiting the outputs of ComE. For computing network level features, we exclude the nodes of the cluster the given node belongs to, to avoid cases of identical network and cluster features. Let $\phi_{v}$ be the embedding of node $v$, and $\phi_{C}$ the embedding of the median of community $C$. Without loss of generality we define our features using the euclidean distance ($d$) between pairs of embeddings. Cosine and $L1$ distances were also evaluated, but euclidean were used as they performed slightly better. In particular, for a node $v$ such that $v\in C$, we define the following features:
\vspace{-0.1in}
\begin{multicols}{2}
\begin{flushleft}
Cluster level:
\vspace{-0.1in}
\begin{itemize}
\item distance from cluster median: $d(\phi_{v}, \phi_{C})$
\item distance from least similar cluster member: $\max_{\forall u \in C} d(\phi_{v}, \phi_u)$
\item distance from most similar cluster member: $\min_{\forall u \in C} d(\phi_{v}, \phi_u)$
\item avg. distance from all cluster members: $avg_{\forall u \in C}$ $d(\phi_{v}, \phi_u)$
\end{itemize}
\end{flushleft}
\vfill\null
\columnbreak
\begin{flushleft}
Network level:
\vspace{-0.1in}
\begin{itemize}
\item min. distance from other cluster median: $\min_{\forall C'\neq C} d(\phi_{v}, \phi_{C'})$
\item distance from least similar node out of the cluster: $\max_{\forall u \notin C} d(\phi_{v}, \phi_u)$
\item distance from most similar node out of the cluster: $\min_{\forall u \notin C} d(\phi_{v}, \phi_u)$
\item avg. distance from all nodes not in the cluster: $avg_{\forall u \notin C}$ $d(\phi_{v}, \phi_u)$
\end{itemize}
\end{flushleft}
\end{multicols}
\vspace{-0.1in}

Our approach is based on the idea that the embeddings of nodes in a cluster are more similar. Thus, if a node has an embedding that is similar to nodes of other clusters, it is more likely to change or leave its cluster. In Fig. \ref{fig4.1:sub2} and Fig. \ref{fig4.1:sub3}, we plot the distribution of the difference, $\Delta$, between $\min_{\forall C'\neq C} d(\phi_{v}, \phi_{C'})$ and $d(\phi_{v}, \phi_{C})$ features, for classes \textit{stay} and \textit{move} for a synthetic dataset generated by Syntgen and a citation DBLP dataset based on \cite{Tang08}. In both figures, we notice that nodes that move to another cluster tend to have lower or even negative $\Delta$ compared to the ones that remain in the same cluster. About 60\% of the \textit{move} nodes have $\Delta$ less than 0.20 for both datasets, while more than 60\% of the \textit{stay} nodes have $\Delta$ higher than 0.28. Therefore, the values of the various distance measures can provide insight on the properties and behavior of a node, and thus, are appropriate as predictive features for our classification problems.

\noindent\textbf{Historical Chains of Features.}
The evolution of a community is tracked in successive network snapshots that correspond to successive timeframes. Thus, all features we describe can be measured for each timeframe. Let us assume a set of $k$ predictive features for each node $v$, and a time period $[1, \dots, n]$, where $f^j_i(v)$ denotes the $j$-th feature of node $v$ at time $i$. Further, for every pair of consecutive timeframes $i$ and $i+1$ in the given time period, the state (label), $l_{i+1}(v)$, of node $v$ can be recorded. 

To model the history of the node and exploit it to derive more accurate predictions, we utilize historical chains of features as defined in \cite{Saganowski15}. In particular, we have as final features for $v$: $\{f^1_1(v), \dots, f^k_1(v)\}$, $l_2(v)$, $\{f^1_2(v),$ $\dots,$ $f^k_2(v)\}$, $l_3(v)$ $\dots$, $\{f^1_n(v),$ $\dots,$ $f^k_n(v)\}$, while $l_{n+1}(v)$ is the label to be predicted. 

Though we have defined here one chain to model the entire node history, the use of subchains of various lengths can also be deployed. While a longer history will provide more information regarding the history of a node, it would limit the number of nodes for which a prediction can be made. 

\begin{table*}
\vspace{-0.25in}
\caption{Snapshot Structure}
\label{tab:dataset}
\centering
\begin{tabular}{c|cc|cc||cc|cc||cc|cc}
\hline
\centering
&\multicolumn{4}{|c||}{\textit{DBLP}}&\multicolumn{4}{|c||}{\textit{Email-eu}}&\multicolumn{4}{|c}{\textit{Syntgen}}\\ \hline
\textbf{Sn$\#$} & \textbf{Nodes} & \textbf{Edges} & \textbf{$|\mathcal{C}|$} & \textbf{$\mathcal{Q}$}& \textbf{Nodes} & \textbf{Edges} & \textbf{$|\mathcal{C}|$} & \textbf{$\mathcal{Q}$}& \textbf{Nodes} & \textbf{Edges} & \textbf{$|\mathcal{C}|$} & \textbf{$\mathcal{Q}$}\\ \hline
0 & 14731 & 120192 & 17 & 0.681& 750 & 4740 & 11 & 0.515 & 1583 & 8955 & 6 & 0.525\\
1 & 16801 & 143404 & 16 & 0.676& 745 & 5077 & 11 & 0.431 & 1443 & 7936 & 5 & 0.480\\
2 & 17756 & 156393 & 16 & 0.649& 742 & 4578 & 10 & 0.528 & 1367 & 7249 & 5 & 0.459\\
3 & 14765 & 120370 & 17 & 0.659& 742 & 4886 & 9 & 0.487 & 1567 & 8154 & 5 & 0.476\\
4 & 10898 & 69005 & 16 & 0.660& 749 & 5072 & 11 & 0.388 & 1575 & 7993 & 5 & 0.483\\
5 &&&&& 739 & 4819 & 10 & 0.521 & 1402 & 6978 & 5 & 0.465\\
6 &&&&& 759 & 4846 & 10 & 0.521 & 1526 & 7629 & 5 & 0.472\\
7 &&&&& 808 & 5405 & 11 & 0.410 & 1416 & 6973 & 5 & 0.462\\
8 &&&&& 772 & 4880 & 13 & 0.391 & 1409 & 6853 & 5 & 0.470\\
9 &&&&& 785 & 5169 & 12 & 0.533 & 1594 & 7794 & 6 & 0.490\\
\hline
\end{tabular}%
\end{table*}
\vspace{-0.25in}
\section{Evaluation}
We experimentally study all three classification problems while comparing different sets of predictive features.

\begin{table*}
\caption{Performance for the $\mathcal{SM}$ Problem}
\label{tab:sm}
\centering
\begin{tabular}{|c|c|c|c|c|c|c|c|c|c|c|}
\hline
\multicolumn{3}{|c|}{}&\multicolumn{4}{|c|}{ComE} &\multicolumn{4}{|c|}{Classic}\\
\hline
\textbf{Data}&\textbf{Feat.}&\textbf{Class} & \textbf{P} & \textbf{R}& \textbf{F1} &\textbf{Acc}& \textbf{P} & \textbf{R}& \textbf{F1} &\textbf{Acc} \\ 
\midrule
\multirow{6}{*}{\textbf{\textit{DBLP}}}&\multirow{2}{*}{\textbf{$in$}} &\textbf{$\mathcal{S}$} &0.821 & 0.940 & 0.876  &\multirow{2}{*}{0.804} &0.837&0.941&0.886&\multirow{2}{*}{0.821}\\
\cline{3-6}\cline{8-10}
&& \textbf{$\mathcal{M}$}& 0.715 & 0.425& 0.533 &&0.746&0.485&0.587&\\
\cline{2-11}
&\multirow{2}{*}{\textbf{$out$}}&\textbf{$\mathcal{S}$} &0.800 & 0.942 & 0.865  &\multirow{2}{*}{0.784}&0.783&0.927&0.849&\multirow{2}{*}{0.757}\\
\cline{3-6}\cline{8-10}
&&\textbf{$\mathcal{M}$}&  0.679 & 0.339& 0.452 &&0.577&0.279&0.376&\\
\cline{2-11}
&\multirow{2}{*}{\textbf{$all$}}&\textbf{$\mathcal{S}$} &0.836& 0.954 &  \textbf{0.891}  &\multirow{2}{*}{0.828}&0.839&0.948&  \textbf{0.890}&\multirow{2}{*}{0.827}\\
\cline{3-6}\cline{8-10}
&&\textbf{$\mathcal{M}$}&  0.787 & 0.476&  \textbf{0.592} &&0.769&0.490& \textbf{0.599}&\\
\midrule
\multirow{6}{*}{\textbf{\textit{Email-eu}}}&\multirow{2}{*}{\textbf{$in$}} &\textbf{$\mathcal{S}$} &0.807 & 0.926 & 0.862  &\multirow{2}{*}{0.796} &0.797&0.912&0.850&\multirow{2}{*}{0.778}\\
\cline{3-6}\cline{8-10}
&& \textbf{$\mathcal{M}$}& 0.757 & 0.507& 0.606 &&0.712&0.483&  \textbf{0.574}&\\
\cline{2-11}
&\multirow{2}{*}{\textbf{$out$}}&\textbf{$\mathcal{S}$} &0.778 & 0.933 & 0.848  &\multirow{2}{*}{0.770}&0.732&0.910&0.812&\multirow{2}{*}{0.709}\\
\cline{3-6}\cline{8-10}
&&\textbf{$\mathcal{M}$}&  0.733 & 0.409& 0.524 &&0.569&0.261&0.357&\\
\cline{2-11}
&\multirow{2}{*}{\textbf{$all$}}&\textbf{$\mathcal{S}$} &0.823& 0.934 &  \textbf{0.875}  &\multirow{2}{*}{0.816}&0.789&0.925&  \textbf{0.852}&\multirow{2}{*}{0.778}\\
\cline{3-6}\cline{8-10}
&&\textbf{$\mathcal{M}$}&  0.792 & 0.554&  \textbf{0.652} &&0.731&0.452&0.558&\\
\midrule
\multirow{6}{*}{\textbf{\textit{Syntgen}}}&\multirow{2}{*}{\textbf{$in$}} &\textbf{$\mathcal{S}$} &0.675 & 0.692 & 0.683  &\multirow{2}{*}{0.668} &0.707&0.719&  \textbf{0.713}&\multirow{2}{*}{0.700}\\
\cline{3-6}\cline{8-10}
&& \textbf{$\mathcal{M}$}& 0.661 & 0.643& 0.652 &&0.693&0.680&0.687&\\
\cline{2-11}
&\multirow{2}{*}{\textbf{$out$}}&\textbf{$\mathcal{S}$} &0.638 & 0.704 & 0.670  &\multirow{2}{*}{0.640}&0.642&0.669&0.655&\multirow{2}{*}{0.635}\\
\cline{3-6}\cline{8-10}
&&\textbf{$\mathcal{M}$}&  0.643 & 0.572& 0.606 &&0.628&0.599&0.613&\\
\cline{2-11}
&\multirow{2}{*}{\textbf{$all$}}&\textbf{$\mathcal{S}$} &0.693& 0.725 &  \textbf{0.708}  &\multirow{2}{*}{0.691}&0.710&0.713&0.711&\multirow{2}{*}{0.701}\\
\cline{3-6}\cline{8-10}
&&\textbf{$\mathcal{M}$}&  0.690 & 0.656&   \textbf{0.672} &&0.691&0.687&  \textbf{0.689}&\\
\midrule
\end{tabular}%
\end{table*}
\newcommand{\bftab}{\fontseries{b}\selectfont}
\begin{table*}
\caption{Performance for the $\mathcal{SL}$ Problem}
\label{tab:sl}
\centering
\begin{tabular}{|c|c|c|c|c|c|c|c|c|c|c|}
\hline
\multicolumn{3}{|c|}{}&\multicolumn{4}{|c|}{ComE} &\multicolumn{4}{|c|}{Classic}\\
\hline
\textbf{Data}&\textbf{Feat.}&\textbf{Class} & \textbf{P} & \textbf{R}& \textbf{F1} &\textbf{Acc}& \textbf{P} & \textbf{R}& \textbf{F1} &\textbf{Acc} \\ 
\midrule
\multirow{6}{*}{\textbf{\textit{DBLP}}}&\multirow{2}{*}{\textbf{$in$}} &\textbf{$\mathcal{S}$} & 0.626 & 0.676 & 0.650  &\multirow{2}{*}{0.922} &0.715&0.687&0.700&\multirow{2}{*}{0.937}\\
\cline{3-6}\cline{8-10}
&& \textbf{$\mathcal{L}$}& 0.961 & 0.951 & 0.956 &&0.962&0.967&0.965&\\
\cline{2-11}
&\multirow{2}{*}{\textbf{$out$}}&\textbf{$\mathcal{S}$} &0.632 & 0.664 & 0.647  &\multirow{2}{*}{0.922}&0.680&0.648&0.664&\multirow{2}{*}{0.929}\\
\cline{3-6}\cline{8-10}
&&\textbf{$\mathcal{L}$}&  0.959 & 0.953& 0.956 &&0.958&0.963&0.960&\\
\cline{2-11}
&\multirow{2}{*}{\textbf{$all$}}&\textbf{$\mathcal{S}$} &0.680& 0.710 & \bftab 0.695  &\multirow{2}{*}{0.933}&0.728&0.690& \bftab 0.709&\multirow{2}{*}{0.939}\\
\cline{3-6}\cline{8-10}
&&\textbf{$\mathcal{L}$}&  0.965 & 0.960& \bftab 0.962 &&0.963&0.969&\bftab 0.966&\\
\midrule
\multirow{6}{*}{\textbf{\textit{Email-eu}}}&\multirow{2}{*}{\textbf{$in$}} &\textbf{$\mathcal{S}$} &0.796 & 0.923 & 0.855  &\multirow{2}{*}{0.820} &0.799&0.892&\bftab 0.843&\multirow{2}{*}{0.809}\\
\cline{3-6}\cline{8-10}
&& \textbf{$\mathcal{L}$}& 0.869 & 0.680& 0.762 &&0.827&0.698&\bftab0.756&\\
\cline{2-11}
&\multirow{2}{*}{\textbf{$out$}}&\textbf{$\mathcal{S}$} &0.771 & 0.915 & 0.837  &\multirow{2}{*}{0.795}&0.725&0.882&0.796&\multirow{2}{*}{0.740}\\
\cline{3-6}\cline{8-10}
&&\textbf{$\mathcal{L}$}&  0.847 & 0.633& 0.724 &&0.775&0.548&0.642&\\
\cline{2-11}
&\multirow{2}{*}{\textbf{$all$}}&\textbf{$\mathcal{S}$} &0.809& 0.934 & \bftab 0.867  &\multirow{2}{*}{0.836}&0.793&0.892&0.840&\multirow{2}{*}{0.804}\\
\cline{3-6}\cline{8-10}
&&\textbf{$\mathcal{L}$}&  0.889 & 0.703& \bftab 0.784 &&0.826&0.686&0.749&\\
\midrule
\multirow{6}{*}{\textbf{\textit{Syntgen}}}&\multirow{2}{*}{\textbf{$in$}} &\textbf{$\mathcal{S}$} &0.656 & 0.654 & 0.655  &\multirow{2}{*}{0.708} &0.696&0.690 & \bftab 0.693&\multirow{2}{*}{0.741}\\
\cline{3-6}\cline{8-10}
&& \textbf{$\mathcal{L}$}& 0.747 & 0.748& 0.747 &&0.774&0.778& \bftab 0.776&\\
\cline{2-11}
&\multirow{2}{*}{\textbf{$out$}}&\textbf{$\mathcal{S}$} &0.634 & 0.660 & 0.646  &\multirow{2}{*}{0.694}&0.631&0.625&0.628&\multirow{2}{*}{0.686}\\
\cline{3-6}\cline{8-10}
&&\textbf{$\mathcal{L}$}&  0.742 & 0.720& 0.731 &&0.726&0.731&0.729&\\
\cline{2-11}
&\multirow{2}{*}{\textbf{$all$}}&\textbf{$\mathcal{S}$} &0.670& 0.682 & \bftab 0.676  &\multirow{2}{*}{0.723}&0.693&0.684&0.688&\multirow{2}{*}{0.738}\\
\cline{3-6}\cline{8-10}
&&\textbf{$\mathcal{L}$}&  0.763 & 0.753& \bftab 0.758 &&0.770&0.778&0.774&\\
\midrule
\end{tabular}%
\end{table*}
\subsection{Datasets}
We use three datasets for our evaluation, two real and one synthetic.

\noindent\textbf{\textit{DBLP:}} \textit{DBLP} is a citation network\footnote{https://www.aminer.org/citation} that includes additional information about the publications, such as year of publication and fields of study they belong to \cite{Tang08}.
Similarly to \cite{ComE17}, we filter papers with primary of study as NLP, Databases, Networking, Data Mining and Computer Vision. We construct five snapshots, for years 2015 to 2019, by maintaining publications of the given year and adding cited papers that belong to the selected fields regardless of their publication year. To attain a denser network, we sample nodes with degree $\geq$ 20 and build the induced undirected subgraph. 

\noindent\textbf{\textit{Email-eu:}} The \textit{Email-eu}\footnote{https://snap.stanford.edu/data/email-Eu-core-temporal.html} \cite{Paranjape17} dataset consists of incoming and outgoing e-mails between members of a large European institution in a period of 803 days. We consider the network undirected and split the data into 10 balanced snapshots. 

\noindent\textbf{\textit{Syntgen:}} To further investigate the impact of network properties on our problems, we use the Syntgen generator \cite{Pereira19} that creates temporal undirected networks simulating real networks using explicit specifications, like degree distributions and cluster sizes, as well as implicitly controlling the perseverance of nodes popularity over time. The intra-cluster to total degree ratio determines cluster density. A high ratio leads to dense well-separated communities, while low values exhibit no clustering. We set the default ratio to 0.7.

We apply ComE \cite{ComE17} to detect communities at each snapshot. To determine an appropriate number of clusters $k$  as input for ComE, for \textit{DBLP} and \textit{Email-eu},  we first apply the Louvain \cite{Blondel08} community detection method that selects the $k$ that maximizes modularity. As to \textit{Syntgen}, we use as $k$ the number of clusters obtained from the generator. Clusters are mapped across different snapshots based on the majority of their common nodes. Table \ref{tab:dataset} presents the number of nodes and edges as well as the number of clusters ($|\mathcal{C}|$) and modularity ($Q$) for each network snapshot for all datasets. 

With regards to historical chains, \textit{DBLP} with 5 snapshots can form 2-length up to 4-length chains, while \textit{Email-eu} and \textit{Syntgen} with 10 snapshots can form from 2-length up to  9-length chains. Trying to balance between more information that longer chains provide and the ability to provide predictions for more nodes, we use as default chain length 5 for both \textit{Email-eu} and \textit{Syntgen}, and 2 for \textit{DBLP}. 

 \begin{figure}
\centering
\begin{subfigure}{0.32\textwidth}
\includegraphics[scale=0.15]{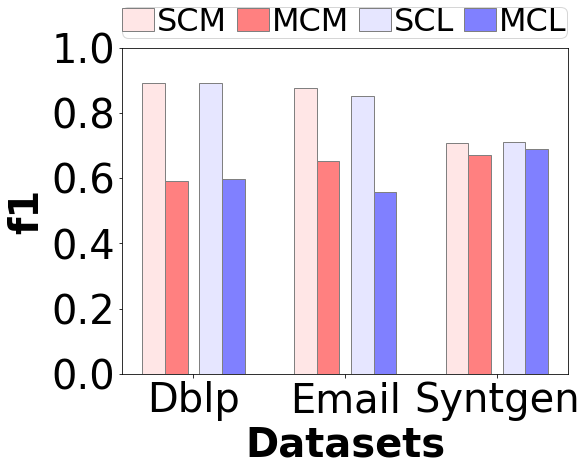}
\caption{$\mathcal{SM}$ Problem}
\label{fig:subSM}
\end{subfigure}
\begin{subfigure}{0.32\textwidth}
\includegraphics[scale=0.15]{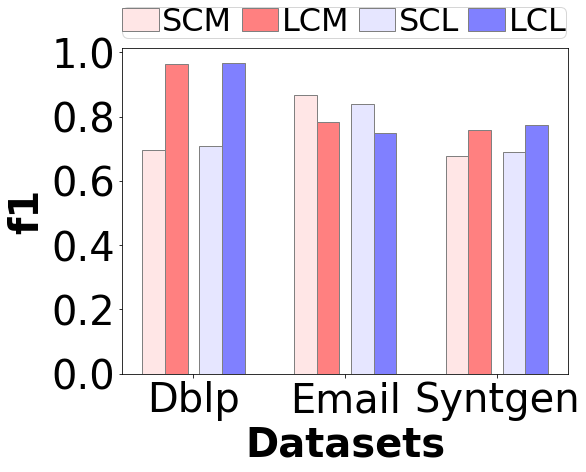}
\caption{$\mathcal{SL}$ Problem}
\label{fig:subSL}
\end{subfigure}
\begin{subfigure}{0.32\textwidth}
\includegraphics[scale=0.15]{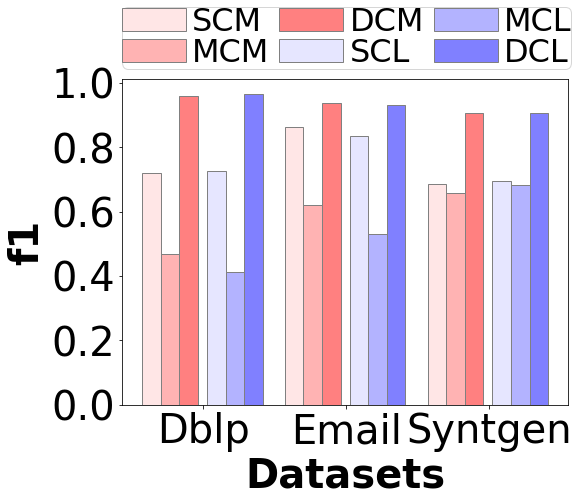}
\caption{$\mathcal{SMD}$ Problem}
\label{fig:subSMD}
\end{subfigure}
\vspace{-0.1in}
\caption{Macro f1 score per class.}
\label{fig:3}
\end{figure}
\vspace{-0.3in}
\subsection{Experimental Results}
We report results using the Random Forest classifier, which performed better compared to other classifiers we tried (e.g. Logistic Regression, Naive Bayes). We apply stratified 5-fold cross-validation to preserve the same class distribution in both train and test sets.
 Tables \ref{tab:sm} to \ref{tab:smd} illustrate precision, recall and f1-score per class, and accuracy for each problem when using ComE and Classic features at cluster ($in$) and network level ($out$), and their combination ($all$) for each dataset. We denote with bold the best f1-score for each class per dataset and problem. As structural features are typically used in community evolution prediction, we consider Classic features as a baseline method to compare the proposed ComE-based features.

\noindent\textbf{General observations.} 
A first observation derived by Table \ref{tab:sm} to Table \ref{tab:smd} is that while the three classification problems are not trivial, both Classic and ComE based features show promising initial results, achieving high accuracy but  showcasing that some classes are more difficult to predict than others. We also notice that $all$ features perform better in most cases, and use them as our default features for the rest of our study.

\begin{table}
\vspace{-0.15in}
\caption{Performance for the $\mathcal{SMD}$ Problem}
\label{tab:smd}
\centering
\begin{tabular}{|c|c|c|c|c|c|c|c|c|c|c|}
\hline
\multicolumn{3}{|c|}{}&\multicolumn{4}{|c|}{ComE} &\multicolumn{4}{|c|}{Classic}\\
\hline
\textbf{Data}&\textbf{Feat.}&\textbf{Class} & \textbf{P} & \textbf{R}& \textbf{F1} &\textbf{Acc}& \textbf{P} & \textbf{R}& \textbf{F1} &\textbf{Acc} \\ 
\midrule
\multirow{9}{*}{\textbf{\textit{DBLP}}}&\multirow{3}{*}{\textbf{$in$}} &\textbf{$\mathcal{S}$} &0.589 & 0.818 & 0.685  &\multirow{3}{*}{0.905} &0.666&0.778&0.718&\multirow{3}{*}{0.917}\\
\cline{3-6}\cline{8-10}
&& \textbf{$\mathcal{M}$}& 0.538 & 0.358& 0.430 &&0.587&0.315&0.410&\\
\cline{3-6}\cline{8-10}
&& \textbf{$\mathcal{D}$}& 0.973 & 0.941& 0.957 &&0.962&0.962&0.962&\\
\cline{2-11}
&\multirow{3}{*}{\textbf{$out$}}&\textbf{$\mathcal{S}$} &0.587 & 0.821 & 0.685  &\multirow{3}{*}{0.904}&0.627&0.744&0.680&\multirow{3}{*}{0.911}\\
\cline{3-6}\cline{8-10}
&&\textbf{$\mathcal{M}$}&  0.491 & 0.250& 0.331 &&0.463&0.161&0.239&\\
\cline{3-6}\cline{8-10}
&& \textbf{$\mathcal{D}$}& 0.972 & 0.944& 0.958 &&0.961&0.966&0.963&\\
\cline{2-11}
&\multirow{3}{*}{\textbf{$all$}}&\textbf{$\mathcal{S}$} &0.621& 0.853 & \bftab 0.718  &\multirow{3}{*}{0.914}&0.682&0.778& \bftab 0.727&\multirow{3}{*}{0.921}\\
\cline{3-6}\cline{8-10}
&&\textbf{$\mathcal{M}$}&  0.601 & 0.384& \bftab 0.468 &&0.616&0.310& \bftab 0.411&\\
\cline{3-6}\cline{8-10}
&& \textbf{$\mathcal{D}$}& 0.976 & 0.945& \bftab 0.960 &&0.963&0.967& \bftab 0.965&\\
\midrule
\multirow{9}{*}{\textbf{\textit{Email-eu}}}&\multirow{3}{*}{\textbf{$in$}} &\textbf{$\mathcal{S}$} &0.795 & 0.932 & 0.858  &\multirow{3}{*}{0.816} &0.783&0.909& \bftab 0.841&\multirow{3}{*}{0.790}\\
\cline{3-6}\cline{8-10}
&& \textbf{$\mathcal{M}$}& 0.748 & 0.517& 0.611 &&0.653&0.468& \bftab 0.545&\\
\cline{3-6}\cline{8-10}
&& \textbf{$\mathcal{D}$}& 1 & 0.881& 0.936 &&0.996&0.881& \bftab 0.935&\\
\cline{2-11}
&\multirow{3}{*}{\textbf{$out$}}&\textbf{$\mathcal{S}$} &0.762 & 0.925 & 0.836  &\multirow{3}{*}{0.785}&0.718&0.897&0.798&\multirow{3}{*}{0.732}\\
\cline{3-6}\cline{8-10}
&&\textbf{$\mathcal{M}$}&  0.686 & 0.413& 0.515 &&0.524&0.272&0.357&\\
\cline{3-6}\cline{8-10}
&& \textbf{$\mathcal{D}$}& 1 & 0.881& 0.936 &&0.993&0.881&0.933&\\
\cline{2-11}
&\multirow{3}{*}{\textbf{$all$}}&\textbf{$\mathcal{S}$} &0.797& 0.939 & \bftab 0.862  &\multirow{3}{*}{0.821}&0.775&0.905&0.835&\multirow{3}{*}{0.784}\\
\cline{3-6}\cline{8-10}
&&\textbf{$\mathcal{M}$}&  0.765 & 0.520& \bftab 0.619 &&0.647&0.451&0.530&\\
\cline{3-6}\cline{8-10}
&& \textbf{$\mathcal{D}$}& 1 & 0.881& \bftab 0.937 &&0.990&0.881&0.932&\\
\midrule
\multirow{9}{*}{\textbf{\textit{Syntgen}}}&\multirow{3}{*}{\textbf{$in$}} &\textbf{$\mathcal{S}$} &0.645 & 0.680 & 0.662  &\multirow{3}{*}{0.693} &0.683&0.707& \bftab 0.695&\multirow{3}{*}{0.724}\\
\cline{3-6}\cline{8-10}
&& \textbf{$\mathcal{M}$}& 0.634 & 0.646& 0.640 &&0.668&0.696&0.681&\\
\cline{3-6}\cline{8-10}
&& \textbf{$\mathcal{D}$}& 1 & 0.828& 0.906 &&1&0.828&0.905&\\
\cline{2-11}
&\multirow{3}{*}{\textbf{$out$}}&\textbf{$\mathcal{S}$} &0.620 & 0.692 & 0.654  &\multirow{3}{*}{0.680}&0.620&0.658&0.638&\multirow{3}{*}{0.670}\\
\cline{3-6}\cline{8-10}
&&\textbf{$\mathcal{M}$}&  0.628 & 0.599& 0.613 &&0.603&0.610&0.606&\\
\cline{3-6}\cline{8-10}
&& \textbf{$\mathcal{D}$}& 1 & 0.828& 0.906 &&1&0.828& \bftab 0.906&\\
\cline{2-11}
&\multirow{3}{*}{\textbf{$all$}}&\textbf{$\mathcal{S}$} &0.661& 0.712 & \bftab 0.685  &\multirow{3}{*}{0.711}&0.685&0.706& \bftab 0.695&\multirow{3}{*}{0.725}\\
\cline{3-6}\cline{8-10}
&&\textbf{$\mathcal{M}$}&  0.659 & 0.657& \bftab 0.658 &&0.668&0.698& \bftab 0.682&\\
\cline{3-6}\cline{8-10}
&& \textbf{$\mathcal{D}$}& 1 & 0.828& \bftab 0.906 &&1&0.828& \bftab 0.906&\\
\midrule
\end{tabular}%
\end{table}

\noindent$\pmb{\mathcal{SM}}$ \textbf{vs.} $\pmb{\mathcal{SL}}$ \textbf{vs.} $\pmb{\mathcal{SMD}}$\textbf{.} To compare the three problems, we illustrate the f1-score for each class for both types of features (ComE, denoted as CM, and Classic denoted as CL) for the three datasets in Fig. \ref{fig:3}. For the $\mathcal{SM}$ problem, 
Fig. \ref{fig:subSM} shows that class \textit{stay} performs better than \textit{move} for \textit{DBLP} and \textit{Email-eu}. \textit{DBLP} achieves the best performance with 0.891 for \textit{stay} and 0.592 for \textit{move} respectively for the ComE features and similar results for the Classic ones (Table \ref{tab:sm}).
 This is due to the imbalance in the real datasets between the two classes, with the majority of the nodes in class \textit{stay}. In contrast, in the \textit{Syntgen} dataset, classes are well-balanced and we notice similar performance for both. 
 For the $\mathcal{SL}$ problem (Fig. \ref{fig:subSL}), we notice  significant difference mainly on \textit{DBLP}. In this case, class \textit{stay} is underrepresented due to the network construction. Looking at Table \ref{tab:sl}, class \textit{stay} achieves f1 0.695 and \textit{leave} 0.962, for \textit{DBLP} with feature type $all$. Finally, as we can see in Table \ref{tab:smd} and Fig. \ref{fig:subSMD} for the $\mathcal{SMD}$ problem, class \textit{move} is heavily underrepresented in both \textit{DBLP} and \textit{Email-eu} resulting in a rather low f1-score. 
 Summing up, the $\mathcal{SL}$ problem seems to have the best overall performance, while $\mathcal{SM}$ appears to have the worst. Apparently, class \textit{move} is overall the most difficult to predict. Besides class imbalance that makes the problem more difficult, this occurs especially when the characteristics across communities are not significantly different, which is a similarity indicator between communities.
\begin{figure}
\centering
\begin{subfigure}{0.32\textwidth}
\includegraphics[scale=0.15]{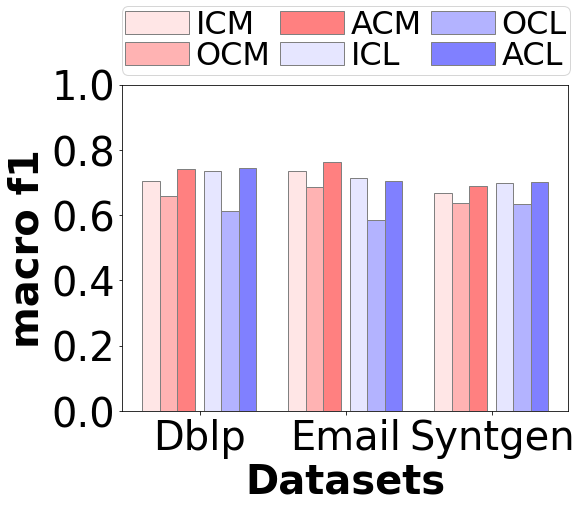}
\caption{$\mathcal{SM}$ Problem}
\label{fig:subim1}
\end{subfigure}
\begin{subfigure}{0.32\textwidth}
\includegraphics[scale=0.15]{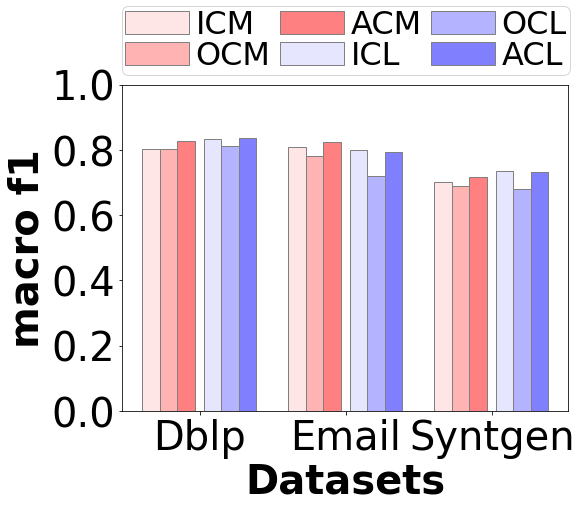}
\caption{$\mathcal{SL}$ Problem}
\label{fig:subim2}
\end{subfigure}
\begin{subfigure}{0.32\textwidth}
\includegraphics[scale=0.15]{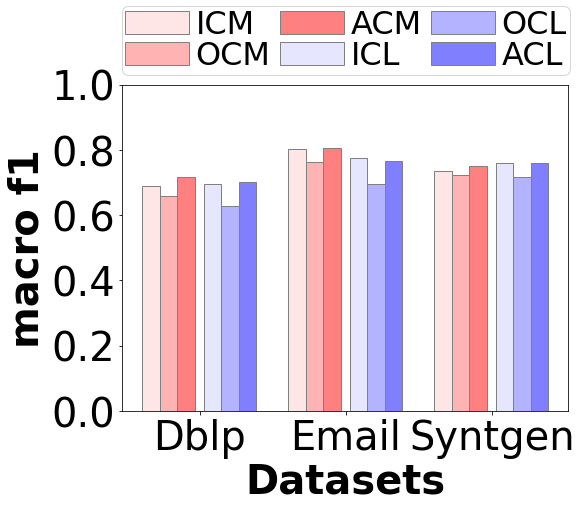}
\caption{$\mathcal{SMD}$ Problem}
\label{fig:subim3}
\end{subfigure}
\vspace{-0.1in}
\caption{Macro f1 score per feature category.}
\label{fig:4}
\end{figure}

\noindent\textbf{Selecting appropriate features.}
Next, we focus on the comparison between the different types of features $in$ (I), $out$ (O) and $all$ (A) for ComE (CM) and Classic (CL) features and how they perform at each problem. 
As we can see in Fig. \ref{fig:4}, $in$ and $all$ outperform $out$ features. For ComE features $all$ is the best choice for all problems, while for Classic features $in$ performs sometimes better. This seems to depend on the dataset and not the problem we study, as we see that for \textit{Email-eu} Classic $in$ features perform better for all problems (with the exception of class \textit{stay} on the $\mathcal{SM}$ problem). 
For the $\mathcal{SM}$ problem and ComE features, we notice a significant difference between $out$ and $all$ features (Fig. \ref{fig:subim1}). Macro f1 is 0.613 for $out$ and 0.744 for $all$ for the \textit{DBLP} dataset and 0.584 for $out$ and 0.704 for $all$ for the \textit{Email-eu} dataset. 
Overall, we do not observe significant differences between ComE and Classic features, deducing that the ComE features offer satisfying performance while being more efficiently computed. For instance, for the \textit{DBLP} dataset, the computation time is 5493s for ComE features, while Classic features are slower by an order, requiring 34618s.

\begin{figure}
\centering
\begin{subfigure}{0.23\textwidth}
\includegraphics[scale=0.18]{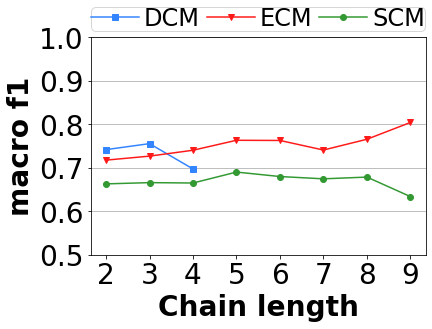}
\caption{$\mathcal{SM}$ Problem}
\label{fig5:subim1}
\end{subfigure}
\begin{subfigure}{0.23\textwidth}
\includegraphics[scale=0.18]{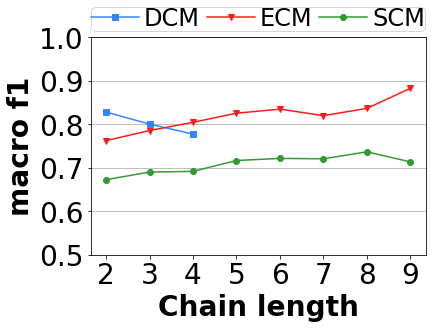}
\caption{$\mathcal{SL}$ Problem}
\label{fig5:subim2}
\end{subfigure}
\begin{subfigure}{0.23\textwidth}
\includegraphics[scale=0.18]{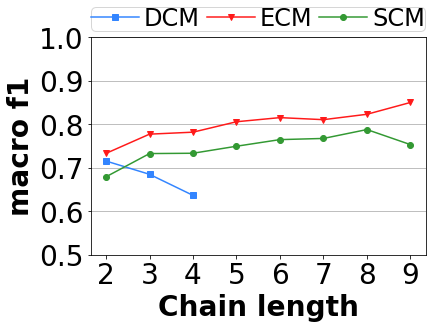}
\caption{$\mathcal{SMD}$ Problem}
\label{fig5:subim3}
\end{subfigure}
\begin{subfigure}{0.23\textwidth}
\includegraphics[scale=0.18]{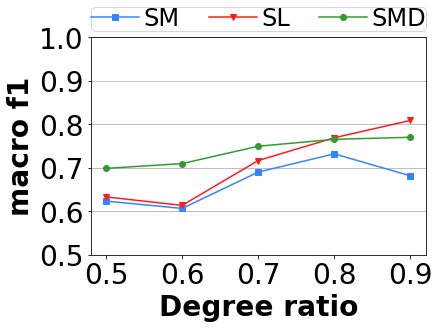}
\caption{Increasing ratio}
\label{fig5:subim4}
\end{subfigure}
\vspace{-0.1in}
\caption{Macro f1 score (a), (b), (c) per chain length and (d) per degree ratio.}
\label{fig:5}
\end{figure}

\noindent\textbf{Influence of the length of historical chains.} Depicted in Fig. \ref{fig5:subim1}, \ref{fig5:subim2}, and \ref{fig5:subim3}, we explore the effect of chains of features with varying length for the \textit{DBLP} (D), \textit{Email-eu} (E) and \textit{Syntgen} (S). Both \textit{Email-eu} and \textit{Syntgen} show that f1 generally improves as chain length grows for all problems. The \textit{Syntgen} dataset follows the same pattern for the $\mathcal{SL}$ and $\mathcal{SMD}$ problems, increasing and reaching its peak at chain length 8. For the \textit{Email-eu}, we observe a temporary drop at length 7, while the highest score is reached at length 9, with 0.804 for the $\mathcal{SM}$ (Fig. \ref{fig5:subim1}), 0.883 for the $\mathcal{SL}$ (Fig. \ref{fig5:subim2}) and 0.850 for the $\mathcal{SMD}$ (Fig. \ref{fig5:subim3}) problem respectively.
As we have mentioned, while longer chains provide more information and more accurate predictions, they are not available for a large number of nodes. In particular, \textit{Email-eu} consists of 6034 instances at 2-length chain and only 750 instances at its longest chain. Similarly, \textit{Syntgen} consists of 11879 instances at 2-length and 1583 instances at 9-length chain. 
With regards to \textit{DBLP}, the best score, 0.828, is achieved at the $\mathcal{SL}$ problem with chain length 2 with 0.828, but the history is too limited to derive safe comparative conclusions.

\noindent\textbf{Influence of intra-cluster to total degree ratio.} In the last experiment, we focus on the Syntgen generator producing different datasets with varying intra-cluster to total degree ratio, which determines the density within the constructed clusters compared to the overall network. In Fig. \ref{fig5:subim4},  we notice a sharp increase on macro f1 for ratio 0.6 up to 0.8 for all problems. Lower ratio indicates poor clustering and thus is not suitable for our context. 
Beyond 0.8, behavior diverges. In such tightly connected communities, most nodes have similar roles in their community, making it difficult for a classifier to determine their behavior.
As a conclusion, cases with ratio close to 0.5 that exhibit no locality, or close to 0.9 indicating almost disconnected communities, fail to achieve good results. 


\vspace{-0.2in}
\section{Conclusions}
In this paper, we defined a novel problem, related to community evolution, that focuses on nodes and aims at predicting whether they will stay in their cluster, move to another or drop out of the network. We modeled the problem as a classification problem and with three variations. We determined appropriate features, based on both local and global node measures, and formed chains of features to take advantage of node history. We also proposed exploiting node and community embeddings derived by the ComE \cite{ComE17} framework to define distance based features. Our experimental results showed that the novel problem is not trivial, and the distance-based features performed similarly to the Classic ones, while requiring far less computation time. Next, we will consider alternative community learning approaches to derive node embeddings and define appropriate features.

\bibliographystyle{splncs04}
\bibliography{newbib}

\end{document}